\documentclass[aps,pre,reprint,superscriptaddress,floatfix]{revtex4-2} % PRE submission (reprint = two-column; use preprint for one-column)
\usepackage[T1]{fontenc}
\usepackage[utf8]{inputenc}
\usepackage{siunitx}
\usepackage{graphicx}
\usepackage{booktabs}
\usepackage{amsmath,amssymb}
\usepackage{color}

\usepackage{flafter}
\begin{document}

\title{An Uncertainty-Aware Equation of State for Gold}

\author{Lin H. Yang} % Write as First name Surname
\email[Corresponding author: ]{lyang@llnl.gov}
\affiliation{Lawrence Livermore National Laboratory, CA 94551, USA}

\author{James A. Gaffney} % Write as First name Surname
\affiliation{Inertia, CA 94550, USA}

\begin{abstract}
We introduce an uncertainty-aware framework for generating equation-of-state (EOS) tables based on Gaussian-process (GP) regression. By incorporating an error-in-variables (EIV) formulation directly into the GP likelihood, the method consistently propagates uncertainties in both the inputs (e.g., density, temperature) and the outputs (e.g., free energies, pressure, and derived thermodynamic quantities) within a single Bayesian framework. In contrast to Monte Carlo ensemble techniques that depend on constructing many EOS tables, a fitted GP delivers, at each state point, a predictive distribution that provides both a mean EOS and uncertainty bands for thermodynamic derivatives. We illustrate the methodology using elemental gold (Au) as a test case, building a first-principles free-energy database from density-functional theory (DFT) that spans mass densities up to \(100~\mathrm{g/cc}\) and temperatures of about \(300~\mathrm{eV}\). Our computational workflow consistently combines cold, electron-thermal, and ion-thermal contributions—employing self-consistent phonon calculations in the solid phase and Born–Oppenheimer molecular dynamics (BOMD)-based modeling in the liquid phase—to generate an uncertainty-quantified equation-of-state (EOS) table (U790). We benchmark this EOS against experimental constraints from diamond-anvil-cell data and shock-Hugoniot measurements and systematically compare it with existing Lawrence Livermore National Laboratory (LLNL) tables (L790 and Y790), enabling us to pinpoint both regions of agreement and the primary sources of disagreement among the models. In addition, we address practical aspects of GP training and outline strategies for extending the approach to larger EOS databases.
\end{abstract}

\maketitle

%===================================================================================
\section{Introduction}
%===================================================================================
High-fidelity equation-of-state (EOS) models are crucial for predicting how materials respond across wide ranges of temperature and pressure. In addition to being accurate and thermodynamically consistent, a practically valuable EOS must remain stable when calibrated on sparse, heterogeneous datasets and should provide a well-justified characterization of uncertainty when applied beyond its training domain. In many current EOS workflows, however, uncertainties in the underlying measurements are not explicitly propagated. Standard tabular approaches such as QEOS/XEOS \cite{More1988,Young1995,Wu2019} typically assume the input data are noise-free and incorporate model error only implicitly, a practice that can systematically underestimate uncertainty in extrapolative settings and encourage overfitting in regions where data are abundant.

A systematic approach for propagating uncertainty in EOS development is to construct ensembles of EOS tables. Ali \textit{et al.} \cite{Ali2020} demonstrated that experimental uncertainties can be robustly propagated through the LLNL multiphase EOS (MEOS) models \cite{Wu2021} and into application-scale simulations via Monte Carlo sampling. Closely related Bayesian calibration and uncertainty quantification methodologies have likewise been developed for multiphase EOS models \cite{LindquistJadrich2022}. While ensemble-based approaches are highly effective, the generation of hundreds to thousands of EOS realizations is frequently computationally prohibitive, thereby motivating the development of representations that encode uncertainty information intrinsically.

Gaussian processes (GPs) constitute a Bayesian, kernel-based regression formalism in which the predictive distribution yields both a mean estimate and an input-dependent characterization of epistemic and aleatoric uncertainty \cite{Rasmussen2005}. 
Within this framework, GPs can serve as smooth surrogate models for thermodynamic potentials and their derivatives, while enabling the incorporation of prior structural information through appropriately chosen covariance kernels and mean functions, including physics-informed constraints derived from thermodynamic consistency requirements \cite{Sharma2024ThermoEOS}. 
Furthermore, GP models can be generalized to accommodate heteroscedastic observation noise, nonstationary behavior, and large-scale datasets by employing sparse or inducing-point approximations \cite{Snelson2006,Titsias2009,QuinoneroRasmussen2005,PaciorekSchervish2006}. 
In particular, the GPz methodology \cite{Almosallam2015SparseGP,Almosallam2016GPz} offers a practical and computationally efficient framework for scalable GP regression with calibrated, and potentially heteroscedastic, predictive uncertainty estimates, and related approaches have been successfully applied in high-energy-density physics for uncertainty-aware inference tasks \cite{Hatfield2019GPzHEDP}.

Standard GP regression typically assumes that the input locations are known exactly and that all uncertainty is confined to the observed outputs. However, EOS datasets frequently violate this assumption: input variables such as densities and temperatures can exhibit substantial uncertainties—originating from experimental calibration errors, finite-size effects, or incomplete numerical convergence—and similar types of errors may correspondingly influence the calculated pressures and free energies.

To account for these effects, we adopt an error-in-variables (EIV) GP framework \cite{Girard2003,Girard2005,McHutchon2011,Johnson2020}, in which uncertainty in the inputs is propagated, via a local Taylor expansion of the latent function, into an effective state-dependent output variance. This construction yields a computationally tractable likelihood function and enables a unified probabilistic treatment of uncertainty in both inputs and outputs. In the limit of vanishing input uncertainty, the EIV–GP formalism recovers standard GP regression as a special case.

We demonstrate this uncertainty-aware equation-of-state (EOS) formulation using gold (Au) as a test case, which serves as a widely adopted reference standard in high-pressure physics. The response of Au at extreme pressures has been extensively characterized by dynamic-compression (shock) experiments \cite{Yokoo2008,Briggs2019,Han2021,colemanBodyCenteredCubicPhaseTransformation2025a} and by static-compression measurements employing diamond-anvil cells \cite{mao2002high,Akahama2002,SHIM2002729,dewaeleEquationsStateSix2004,dubrovinskaiaTerapascalStaticPressure2016a,Dewaele2018,Dewaele2007,Takemura2008}. Complementary first-principles calculations within the DFT framework have yielded detailed predictions for the behavior of Au under compression and at elevated temperatures \cite{Soderlind2002,dubrovinskyNoblestAllMetals2007,ishikawaPressureinducedStackingSequence2013,ahujaTheoreticalPredictionPhase2001,boettgerTheoreticalExtensionGold2003,Liu2016,Smirnov2017,richardInitioPhaseDiagram2023,colemanBodyCenteredCubicPhaseTransformation2025a}; however, substantial differences persist in the high-pressure regime. Therefore, it is crucial to rigorously characterize these uncertainties and monitor how they propagate in order to develop EOS tables with defensible and clearly documented credibility bounds. In the LLNL EOS table nomenclature for Au (material identifier 790), we refer to the existing tables as L790 and Y790, and we label the newly developed uncertainty-aware table as U790.

The remainder of this paper is organized as follows. 
Section~\ref{sec:GP} summarizes the EIV--GP formulation and its role in EOS table construction, including how input uncertainties can be included when available.
Section~\ref{sec:F_DFT} describes the first-principles free-energy components used for Au.
Section~\ref{sec:DFTuncertainty} discusses major uncertainty sources in the DFT database and how they are mapped into the GP training noise model.
Section~\ref{sec:UEOS} presents the resulting uncertainty-aware Au EOS table and comparisons with established tables and experiments. 
We conclude in Sec.~\ref{sec:conclusions} with a discussion of scalability and downstream use in uncertainty propagation.

\section{Gaussian processes for EOS with input and output uncertainty}
\label{sec:GP}

GP regression provides a highly flexible surrogate model for thermodynamic potentials and inherently furnishes a full predictive distribution—comprising both mean and variance—at each state point \cite{Rasmussen2005}. 
In UEOS, we employ an EIV--GP framework to rigorously propagate uncertainties originating from noisy training data—including uncertainties in both the input state variables and the observed thermodynamic quantities—into a smooth, differentiable representation that is well suited for the construction of EOS tables.

\subsection{Model definition and notation}

We consider a latent function $f(\mathbf{X})$ mapping a thermodynamic state $\mathbf{X}$ to an observable (e.g., a free-energy component). 
The input dimension $D$ depends on the quantity being modeled. For example, the cold curve uses $D=1$ with $\mathbf{X}=\rho$, whereas thermal contributions use $D=2$ with $\mathbf{X}=(\rho,T)$.
For a generic training set $\{(\mathbf{X}_i,y_i)\}_{i=1}^{N}$ we assume
\begin{align}
y_i &= f(\tilde{\mathbf{X}}_i) + \epsilon_{y,i}, \qquad \epsilon_{y,i}\sim\mathcal{N}(0,\sigma_{y,i}^{2}), \\
\mathbf{X}_i &= \tilde{\mathbf{X}}_i + \boldsymbol{\epsilon}_{\mathbf{X},i}, \qquad \boldsymbol{\epsilon}_{\mathbf{X},i}\sim\mathcal{N}(\mathbf{0},\Sigma_{\mathbf{X},i}),
\end{align}
where $\tilde{\mathbf{X}}_i$ are latent (true) inputs, $\sigma_{y,i}$ are output uncertainties, and $\Sigma_{\mathbf{X},i}$ are input-noise covariances.
When $\Sigma_{\mathbf{X},i}=\mathbf{0}$ the formulation reduces to the standard noisy-output GP.

In this work we construct the total Helmholtz free energy by modeling its components with separate GPs and then combining them algebraically,
\begin{equation}
F(\rho,T)=F_{\mathrm{cold}}(\rho)+F_{\mathrm{vib}}(\rho,T)+F_{\mathrm{elec}}(\rho,T).
\label{eq:F_decomp}
\end{equation}
Each constituent GP model is trained using an input dataset $\mathbf{X}$ and corresponding outputs $y$. The output uncertainties are prescribed according to the DFT and experimental uncertainty characterization described in Sec.~\ref{sec:DFTuncertainty}. For the DFT datasets, the $(\rho, T)$ state points are fixed by the simulation grid, and we therefore set $\Sigma_{\mathbf{X},i} = \mathbf{0}$. In contrast, the EIV framework becomes crucial when incorporating experimental datasets, in which state variables (e.g., density and temperature) are themselves uncertain, and when propagating this state-point uncertainty to derived thermodynamic quantities.

Accordingly, the cold-curve GP uses $\mathbf{X}=\rho$ and $y=F_{\mathrm{cold}}$, whereas the electron-thermal and ion-thermal GPs use $\mathbf{X}=(\rho,T)$ with outputs $F_{\mathrm{elec}}$ and $F_{\mathrm{vib}}$, respectively. For all three models, the output-noise specification is assigned from Table~\ref{tab:Au_DFTuncertainty}, while $\Sigma_{\mathbf{X},i}=\mathbf{0}$ for the present DFT datasets.

In the UEOS framework, we make use of stationary kernels equipped with automatic relevance determination (ARD). By default, we adopt the squared-exponential (SE) kernel,
\begin{equation}
k_{\mathrm{SE}}(\mathbf{X},\mathbf{X}')=\sigma_f^2\exp\!\left[-\frac12(\mathbf{X}-\mathbf{X}')^{\mathsf{T}}\Lambda^{-1}(\mathbf{X}-\mathbf{X}')\right],
\label{eq:k_se}
\end{equation}
where $\Lambda=\mathrm{diag}(\lambda_1^2,\ldots,\lambda_D^2)$ and $\theta=\{\sigma_f,\lambda_1,\ldots,\lambda_D\}$ are kernel hyperparameters optimized by marginal-likelihood maximization (Appendix~\ref{app:eivgp}).

To assess sensitivity to the assumed smoothness---and, consequently, to the inferred local curvature of the free-energy surface near ambient conditions---we also consider Mat\'ern kernels with fixed $\nu=3/2$ and $\nu=5/2$,
\begin{align}
k_{3/2}(\mathbf{X},\mathbf{X}')
&=\sigma_f^{2}\left(1+\sqrt{3}\,r\right)\exp\!\left(-\sqrt{3}\,r\right),
\label{eq:k_m32}\\
k_{5/2}(\mathbf{X},\mathbf{X}')
&=\sigma_f^{2}\left(1+\sqrt{5}\,r+\frac{5}{3}r^{2}\right)\exp\!\left(-\sqrt{5}\,r\right),
\label{eq:k_m52}
\end{align}
with the ARD distance
\begin{equation}
r=\left[(\mathbf{X}-\mathbf{X}')^{\mathsf{T}}\Lambda^{-1}(\mathbf{X}-\mathbf{X}')\right]^{1/2}.
\label{eq:ard_dist}
\end{equation}
The SE kernel is infinitely differentiable, whereas the Mat\'ern-$3/2$ and Mat\'ern-$5/2$ kernels enforce only finite degrees of smoothness, corresponding to well-defined first- and second-order thermodynamic derivatives, respectively. All of these covariance functions possess closed-form analytical derivatives, enabling uncertainty propagation through thermodynamic derivatives of the fitted free-energy surfaces (see Appendix~\ref{app:eivgp} for explicit expressions of the Mat\'ern derivatives).

\subsection{Prior mean functions and physics-informed structure}

A GP model can be expressed as
\[
f(\mathbf{X}) = m(\mathbf{X}) + g(\mathbf{X}),
\]
where \(m(\mathbf{X})\) denotes a specified mean function and \(g(\mathbf{X})\) is a zero-mean GP.  
Incorporating a physics-informed mean function—such as a Vinet cold curve \cite{Vinet1987} or a baseline QEOS/XEOS model \cite{More1988,Young1995}—effectively allocates the GP component to modeling the residual, thereby reducing extrapolative variance.  
In this study, we employ a zero-mean GP by standardizing each dataset; however, the UEOS framework can also accommodate nonzero mean functions, which provides a powerful way to curb variance inflation, especially in sparsely sampled regions of the input space.

\subsection{Including input uncertainty: error-in-variables likelihood}

The EIV–GP framework \cite{Girard2003,Girard2005,McHutchon2011,Johnson2020} incorporates input uncertainty by performing a local (usually first-order) expansion of the latent function and then propagating the input noise, resulting in an effective output variance that is generally heteroscedastic.
\begin{equation}
\sigma_{\mathrm{eff},i}^{2}\approx\sigma_{y,i}^{2}+\nabla f(\tilde{\mathbf{X}}_i)^{\mathsf{T}}\Sigma_{\mathbf{X},i}\nabla f(\tilde{\mathbf{X}}_i).
\end{equation}
In the special case where $\Sigma_{\mathbf{X},i}=\mathbf{0}$, this additional contribution vanishes. Since $\sigma_{\mathrm{eff},i}^{2}$ depends on the posterior gradient of the latent function, training in the EIV framework is carried out via an iterative procedure that alternates between updating $\Sigma_{\mathrm{eff}}$ and optimizing the kernel hyperparameters (Appendix~\ref{app:eivgp}).

\begin{figure}[htbp]
    \centering
        \includegraphics[width=\columnwidth, trim=0cm 0cm 0cm 0cm,clip]{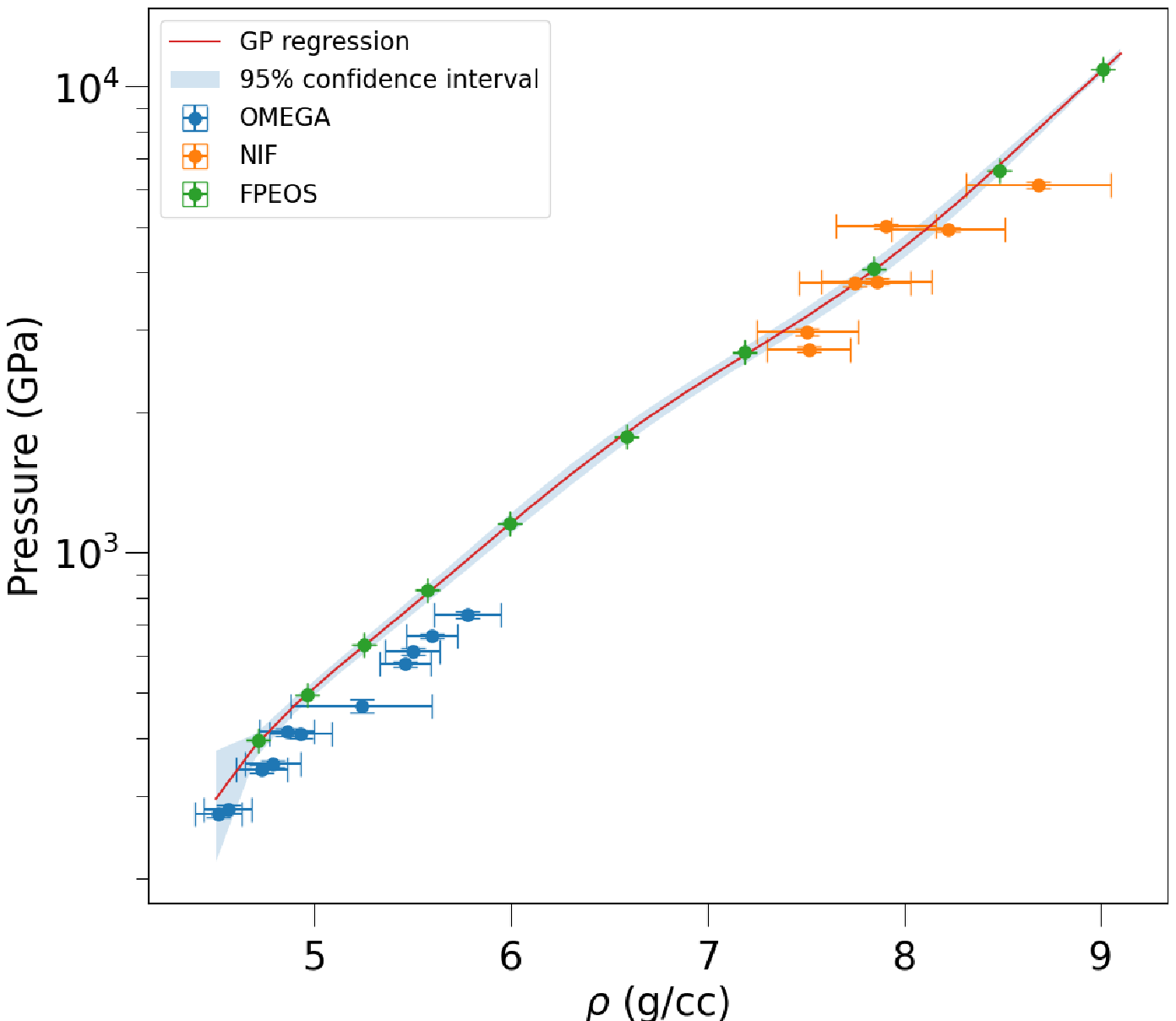}

    \caption{Illustration of EIV--GP regression incorporating uncertainties in both the input and output domains. The figure presents the predicted mean and associated credible interval of the B$_4$C principal Hugoniot obtained by combining theoretical equation-of-state data (FPEOS) \cite{Militzer2021} with experimental measurements \cite{Zhang2020}. }
    \label{fig:B4C_Hugoniot}
\end{figure}

Figure~\ref{fig:B4C_Hugoniot} illustrates how UEOS incorporates measurement uncertainties, when available. The experimental Hugoniot data points include uncertainties in density (represented by horizontal error bars), which are treated as noise in the input variables for the regression, while pressure uncertainties (vertical error bars) are modeled as noise in the output variable \cite{Zhang2020,Militzer2021}. Thus, horizontal error bars are interpreted as uncertainty in the predictor (density), whereas vertical error bars are interpreted as observational uncertainty in the response (pressure).

For datasets that contain several levels of fidelity (such as low-fidelity simulations and high-fidelity experimental observations), this regression framework can be extended to a multi-fidelity GP (co-kriging) approach, which jointly learns correlations across fidelities as well as systematic discrepancies between them. In this setting, low-fidelity data primarily shape the broad, global structure of the surrogate model, while high-fidelity data deliver targeted refinement and calibration of the regression, reducing the chance that any single fidelity level dominates the overall model fit. 

This feature is especially crucial for combining experimental and computational datasets, as it provides a systematic way to propagate uncertainties in state points into thermodynamic properties obtained from the EOS.

\begin{figure}[htbp]
    \centering
        \includegraphics[width=\columnwidth, trim=0cm 0cm 0cm 0cm,clip]{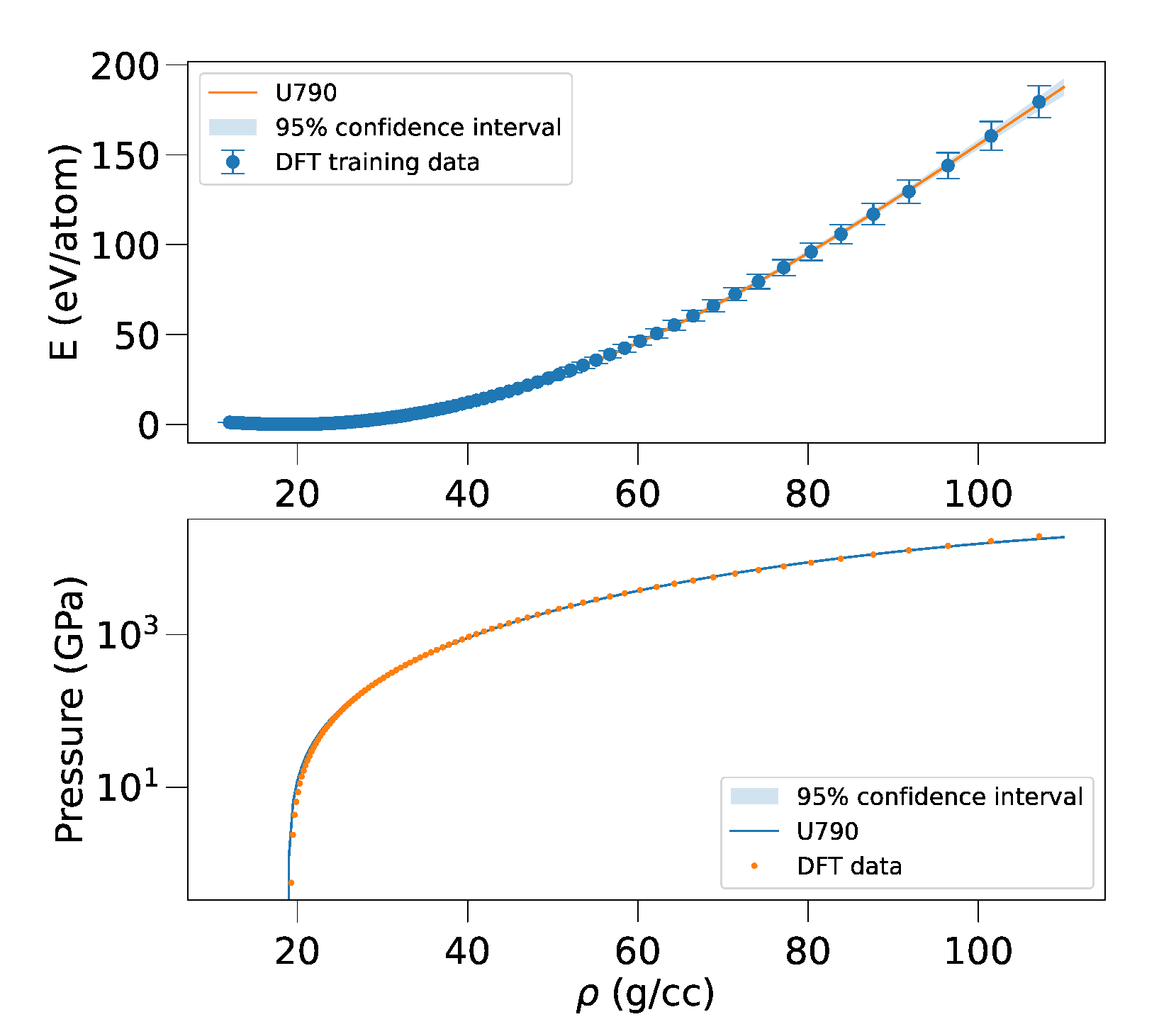}

    \caption{UEOS regression for the gold (Au) cold curve. DFT data are used to train an EIV--GP; in this case $\Sigma_{\mathbf{X}}=\mathbf{0}$, which enables analytic evaluation of $\partial F_{\mathrm{cold}}/\partial \rho$ and thereby facilitates the propagation of the model’s predictive uncertainty to the pressure. The Hellmann--Feynman pressure values obtained from the DFT calculations are represented as dotted data points.}
    \label{fig:Au_cold_EOS}
\end{figure}

GPs are nonparametric models that can capture smooth, nonlinear relationships without requiring a predetermined analytic functional form. Figure~\ref{fig:Au_cold_EOS} demonstrates how the UEOS framework is used to fit the DFT cold curve and to transfer the resulting uncertainty to the pressure through analytic differentiation of the fitted energy.

%===================================================================================
\section{First-principles Helmholtz free energy}
\label{sec:F_DFT}
%===================================================================================
We examine the first-principles free-energy contributions that form the basis of the Au EOS. In this study, we omit explicit electron–phonon coupling at finite temperatures. Instead, we decompose the free energy into ground-state (cold), electron-thermal, and ion-thermal components. Each contribution characterizes distinct aspects of the system’s behavior across different temperature and pressure regimes. To obtain quantitatively reliable energy values, we employed plane-wave pseudopotential methodologies within the DFT framework \cite{Yang2007,Giannozzi2009}. The exchange–correlation energy was treated using several functionals, namely the local density approximation (LDA) \cite{perdewSelfinteractionCorrectionDensityfunctional1981}, the Perdew–Burke–Ernzerhof (PBE) generalized gradient approximation (GGA) \cite{Perdew1996,perdewGeneralizedGradientApproximation1997}, and the PBE functional optimized for solids (PBEsol) \cite{Terentjev2018}. In addition, relativistic effects were incorporated through appropriate treatments \cite{dalcorsoProjectorAugmentedwaveMethod2010,dalcorsoProjectorAugmentedWave2012}, which are essential for achieving a comprehensive and accurate description of the electronic structure and associated energetic properties. The systematic use of multiple exchange–correlation functionals within DFT also facilitates an assessment of methodological uncertainties, thereby enabling a more robust characterization of the system’s response over a range of temperatures and pressures.

Low-temperature liquid states were investigated using Born–Oppenheimer molecular dynamics (BOMD) simulations. To achieve a more accurate representation of anharmonic lattice dynamics, we additionally applied the self-consistent phonon framework to quantitatively assess the influence of temperature-dependent vibrational contributions. This methodology, rigorously developed by Tadano \textit{et al.} \cite{Tadano2015}, enables a precise treatment of anharmonic effects on both thermodynamic and dynamical properties.

To faithfully capture the broad range of mass densities (\(0.5\rho_0\)–\(5\rho_{0}\)) and temperatures (up to 300 eV) required for the Au EOS table, we constructed 
an optimized norm-conserving Vanderbilt pseudopotential (ONCVpsp) \cite{Hamann2013,Pask2023}. The resulting Au ONCVpsp treats 33 electrons 
as valence, corresponding to the \(4f^{14}\), \(5s^{2}\), \(5p^{6}\), \(5d^{10}\), and \(6s^{1}\) shells, and uses a local potential cutoff radius of \(r_c = 1.1~\text{bohr}\). This configuration provides high accuracy together with computational efficiency for the DFT calculations that form the basis of the Au EOS. We verified the robustness of this pseudopotential by comparing computed bulk properties with 
existing theoretical and experimental results, as reported in Table~\ref{tab:Au_bulk}.
The inclusion of an extensive set of valence states improves the pseudopotential’s transferability under extreme pressures and temperatures, where significant electronic excitation takes place. 
In addition, to achieve self-consistent convergence of the total energy to better than 1 meV per atom and pressure convergence within a few kbar, we adopted a plane-wave kinetic-energy cutoff of 175 Hartree and a Brillouin-zone \(k\)-point mesh corresponding to a maximum spacing of \(0.2~\text{\AA}^{-1}\).

\begin{table}[t]
\centering
\caption{Equilibrium properties of fcc Au predicted by several exchange--correlation functionals. The PBEsol and PBEsol+S--O results are in the closest agreement with the reference data listed in Table~\ref{tab:Au_bulk}.}
\label{tab:Au_DFT}
\begin{tabular}{lcccc}
\toprule
XC functional & $V_0$ (\AA{}$^3$) & $\rho_0$ (g/cc) & $B_0$ (GPa) & $B_0'$ \\
\midrule
LDA           & 16.57 & 19.72 & 194 & 5.61 \\
PBE           & 17.96 & 18.20 & 139 & 5.83 \\
PBEsol        & 17.00 & 19.22 & 175 & 5.90 \\
LDA+S--O      & 16.46 & 19.85 & 199 & 5.78 \\
PBE+S--O      & 17.81 & 18.35 & 144 & 5.83 \\
PBEsol+S--O   & 16.92 & 19.31 & 179 & 5.88 \\
\bottomrule
\end{tabular}
\end{table}

\begin{table}[t]
\centering
\caption{Reference equilibrium properties of fcc Au used to benchmark the present DFT calculations. The values reported here are most closely matched by the PBEsol results in Table~\ref{tab:Au_DFT}.}
\label{tab:Au_bulk}
\begin{tabular}{lccc}
\toprule
Method & $V_0$ (\AA{}$^3$) & $B_0$ (GPa) & $B_0'$ \\
\midrule
This work (ONCVpsp)          & 17.00 & 175 & 5.90 \\
FP-LMTO \cite{Smirnov2017}   & 17.10 & 171 & 5.78 \\
Exp. (300 K) \cite{Dewaele2007} & 16.96 & 167 & 5.88 \\
Exp. (0 K) \cite{Neighbours1958} & -- & 180 & -- \\
\bottomrule
\end{tabular}
\end{table}

The Helmholtz free energy \((F = U - TS)\) for each phase can be expressed as the sum of three distinct contributions:
\[
F(\rho, T) = F_{\mathrm{cold}}(\rho) + F_{\mathrm{vib}}(\rho, T) + F_{\mathrm{elec}}(\rho, T),
\]
where \(F_{\mathrm{cold}}(\rho)\) denotes the cold-curve energy, i.e., the internal energy of the crystalline phase at \(T = 0\,\mathrm{K}\); \(F_{\mathrm{vib}}(\rho, T)\) represents the ion-thermal contribution arising from the thermal vibrations of ions in both solid and liquid states; and \(F_{\mathrm{elec}}(\rho, T)\) corresponds to the electron-thermal contribution associated with thermally excited electrons, typically modeled via Fermi–Dirac smearing.  

In this work, we focus on the solid fcc phase and the liquid state. This narrower scope is intended as a methodological demonstration of the UEOS framework; a complete multiphase EOS for Au, together with the associated thermodynamic properties, will be reported separately \cite{Yang2026}.

This section outlines the methodology employed to derive the Helmholtz free energy from DFT calculations.

\subsection{Ground state energy}

\begin{figure}[htbp]
    \centering
        \includegraphics[width=\columnwidth, trim=0cm 0cm 0cm 0cm,clip]
        {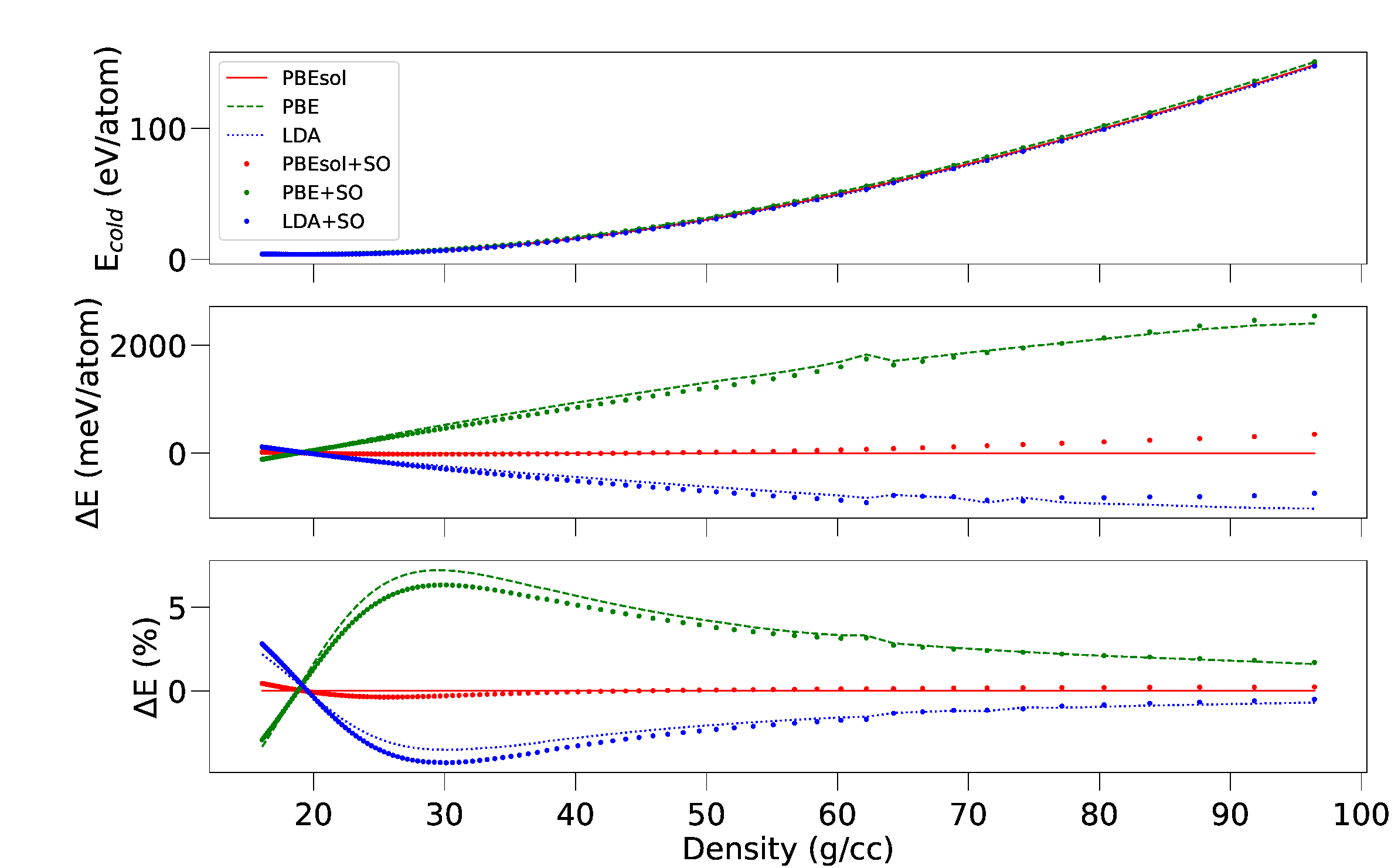}

    \caption{Comparison of cold EOSs for fcc Au using different density functional forms and S--O coupling. For the energy difference, we use PBEsol as the reference.   }
    \label{fig:DFT-cold_UQ_fcc}
\end{figure}

The ground-state energies of the fcc phase were calculated using DFT within the plane-wave pseudopotential framework. Several exchange–correlation functionals were applied, including LDA, PBE, PBEsol, as well as functionals that account for relativistic effects. The resulting equilibrium bulk properties of Au from these different DFT approaches are listed in Table~\ref{tab:Au_DFT}. Among the tested functionals, PBEsol and PBEsol with spin–orbit (S–O) coupling exhibit the closest agreement with the reference bulk data given in Table~\ref{tab:Au_bulk}. Figure~\ref{fig:DFT-cold_UQ_fcc} shows the corresponding cold curves and their deviations from the PBEsol reference.

\begin{figure}[htbp]
    \centering
        \includegraphics[width=\columnwidth, trim=0cm 0cm 0cm 0cm,clip]{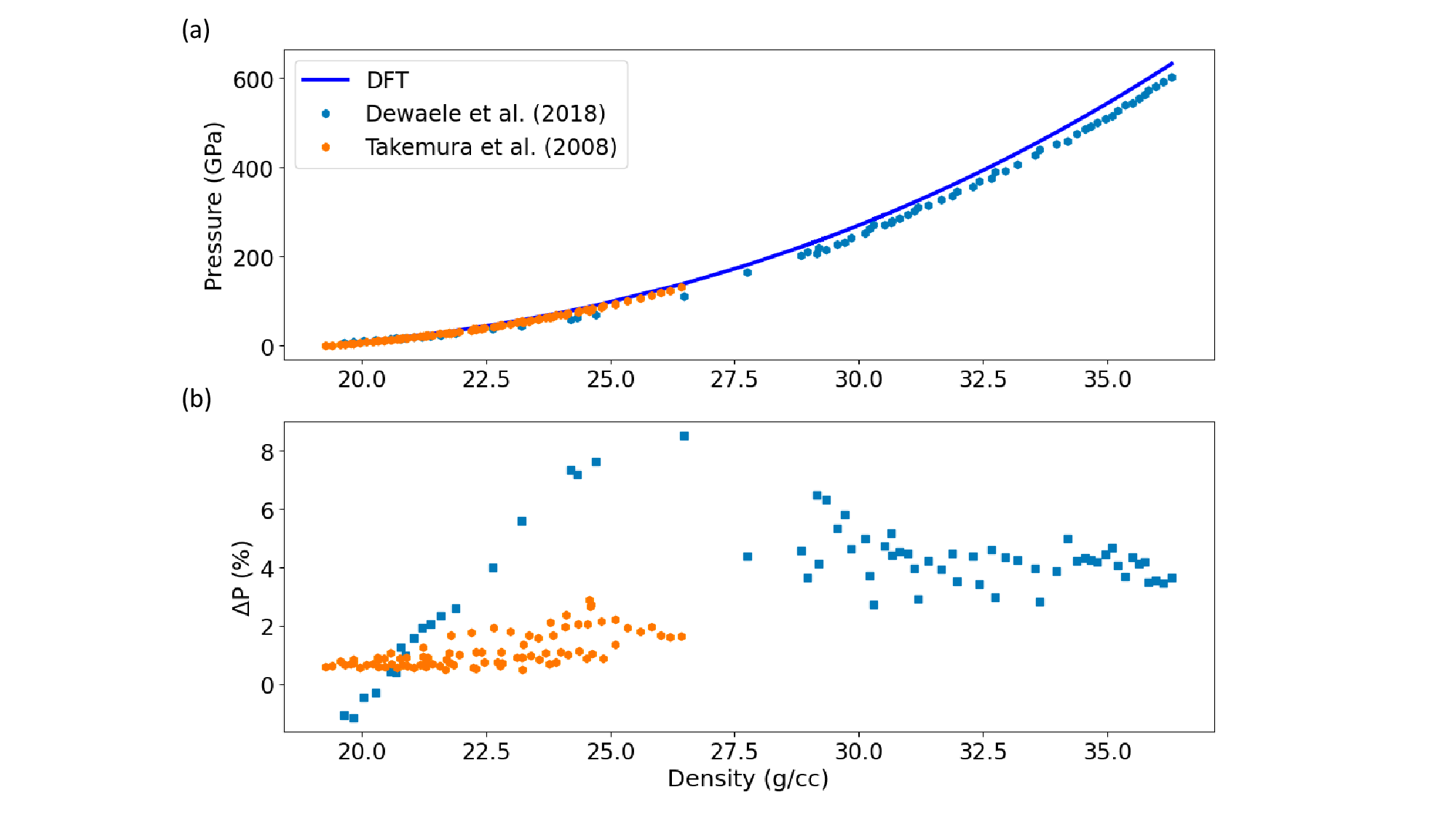}

    \caption{(a) Au isotherm data from DFT calculations (solid line), DAC \cite{Takemura2008} (orange hexagons), and tDAC  \cite{Dewaele2018} (blue squares) measurements. (b) The percent difference between DFT isotherm data and experiments.}
    \label{fig:DFT-DAC}
\end{figure}

For the high-pressure fcc phase, we performed a detailed comparison of isothermal pressure as a function of density using both our DFT calculations and diamond anvil cell (DAC) measurements \cite{Takemura2008}, spanning densities from 19.28 ($\rho_{0}$) to 26.5 g/cc. In this analysis, we examined the pressure differences between our DFT results and the DAC data. The two datasets agree very well overall, with deviations remaining below 2.5\%, as illustrated in Fig.~\ref{fig:DFT-DAC} (b). We then extended this comparison to higher pressures, up to 600 GPa (36 g/cc), by comparing our DFT results with recent toroidal diamond anvil cell (tDAC) measurements by Dewaele \textit{et al.} \cite{Dewaele2018}. Under these more extreme conditions, the discrepancies between the DFT and tDAC data become more pronounced, reaching up to about 8\% near 27 g/cc.

\subsection{Electron-thermal free energy}

Following a strategy similar to that used in ground-state solid-state calculations, we obtained the electron-thermal free energy by solving the Kohn–Sham equations and applying Fermi–Dirac statistics to describe electron occupations at finite temperature. As the ionic reference configuration, we chose the fcc solid at T = 0 K. At each point on the density (\(\rho\))–temperature (T) grid, we self-consistently evaluated both the internal energy, \( U_{\mathrm{elec}}(\rho, T) \), and the free energy, \( F_{\mathrm{elec}}(\rho, T) \). Furthermore, we utilized snapshots from BOMD simulations as additional ionic reference configurations. This approach allowed us to quantify the uncertainties in the electron-thermal free energy, especially those stemming from changes in electronic occupation at nonzero temperatures.

\subsection{Ion-thermal free energy}

To calculate the vibrational contributions to the Helmholtz free energy \(F_{\mathrm{vib}}\) and the internal energy \(U_{\mathrm{vib}}\), we employed self-consistent phonon (SCP) calculations. SCP theory provides a nonperturbative framework for evaluating temperature-dependent phonon frequencies. In this approach, an effective harmonic phonon frequency and polarization vector are introduced, which define an effective harmonic Hamiltonian. The renormalized phonon frequencies and eigenvectors are then obtained by minimizing the vibrational free energy within the first-order cumulant approximation.

In SCP theory, an anharmonic crystal is modeled by a Hamiltonian composed of a harmonic term augmented by anharmonic contributions of progressively higher order. In its conventional formulation, SCP theory accounts for quartic anharmonicity while neglecting cubic anharmonic terms. By contrast, the improved self-consistent (ISC) phonon theory generalizes this formulation by explicitly incorporating an additional three-phonon interaction term, which is treated within perturbation theory. The resulting correction to the Helmholtz free energy is represented diagrammatically by the bubble diagram corresponding to cubic anharmonicity \cite{Tadano2015}.

\paragraph{Vibrational free energies for solids}

Within the SCP framework, the second-, third-, and fourth-order interatomic force constants (IFCs) are obtained from first-principles calculations. This is achieved by systematically displacing atoms in a supercell and calculating the forces generated by these displacements. The resulting force–displacement data set is then fitted using the Compressive Sensing Lattice Dynamics (CSLD) method \cite{Zhou2019,Tadano2015} to extract the IFCs with high accuracy. For fcc Au, we employed 65 unique atomic displacement configurations in a 64-atom supercell at each density to determine these constants. To assess finite-size effects on the calculated energy, we additionally carried out simulations with 108-atom and 256-atom supercells. Across the full density range considered in this study, the deviation remains on average below \( 1\% \). The uncertainty associated with the choice of supercell size will be discussed in Sec.\ref{sec:DFTuncertainty}.

The SCP equation is solved using numerical optimization schemes, in particular the least absolute shrinkage and selection operator (LASSO) method. The SCP \( \mathbf{q} \)-point mesh of \( 4\times 4\times 4 \) is chosen to be commensurate with the supercell size, and an inner \( \mathbf{q} \)-point mesh is systematically refined to achieve convergence of the anharmonic phonon frequencies. The resulting effective dynamical matrices are transformed into real-space effective second-order interatomic force constants (IFCs), which are then employed to compute anharmonic phonon frequencies on a substantially denser \( \mathbf{q} \)-point grid. For fcc Au, a \( 51\times 51\times 51 \) \( \mathbf{q} \)-point mesh is used. In comparison with Born–Oppenheimer molecular dynamics (BOMD), the use of SCP theory to evaluate the ion-thermal contribution to the free energy provides a significant advantage, as it rigorously preserves the crystalline symmetry of the solid. At the same time, SCP theory enables an efficient and accurate determination of the vibrational free energy.

\paragraph{Vibrational free energies for liquids}

\begin{figure}[htbp]
    \centering
        \includegraphics[width=\columnwidth, trim=0cm 0cm 0cm 0cm,clip]{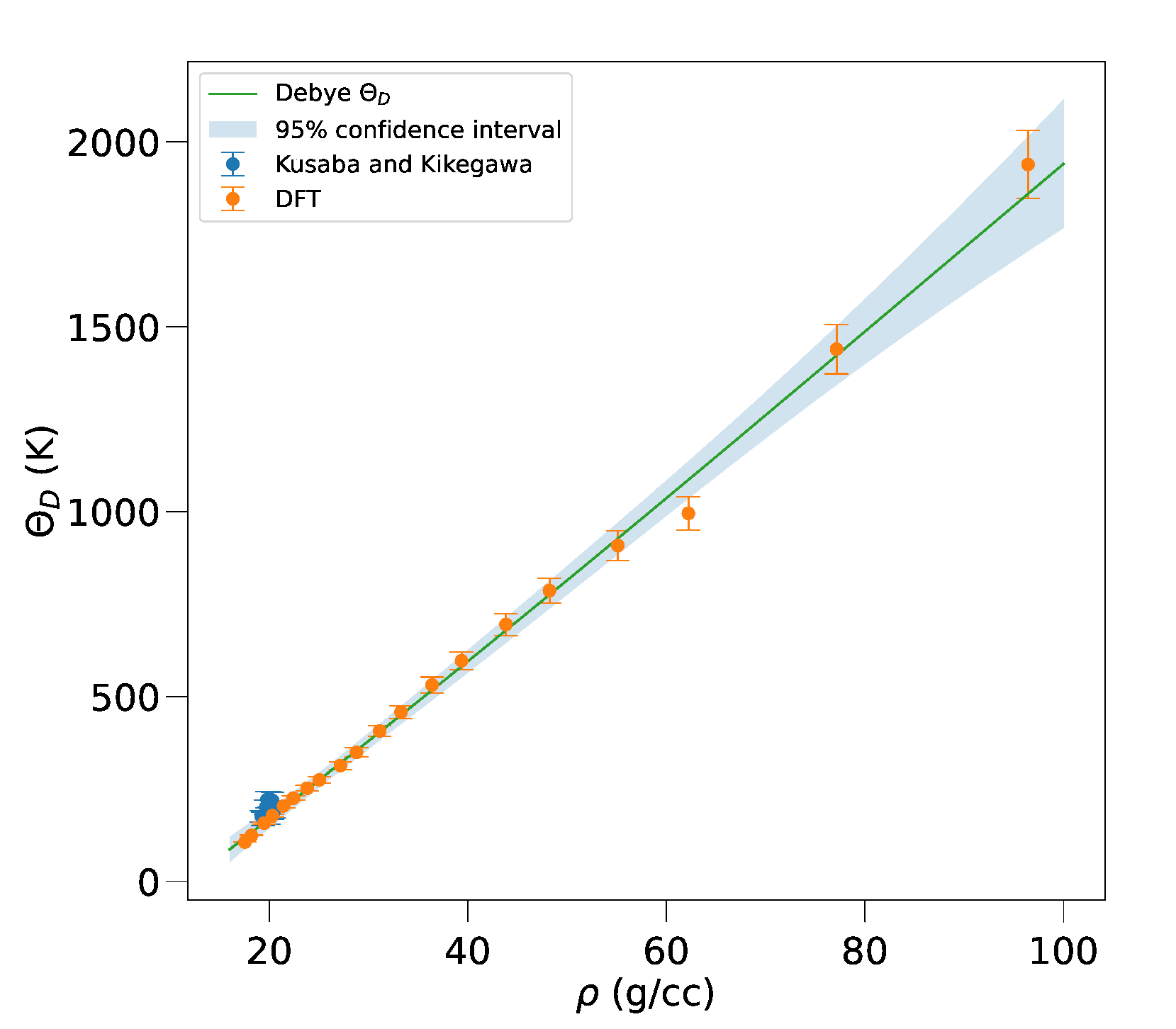}

    \caption{Debye temperature of Au as a function of density from DFT calculations (orange dots) and experimental data \cite{KUSABA2009371}. The blue solid line represents a GP fit, with the 95\% credible interval indicated by the shaded region.}
    \label{fig:DFT-Debye}
\end{figure}

For low-temperature liquids, our SCP calculations rely on configurations sampled from BOMD simulations. Using these configurations, we obtained finite-temperature IFCs and phonon frequencies via the SCP approach, as previously described for the solid phase. To extend the SCP internal-energy and free-energy descriptions to higher temperatures within the warm dense matter (WDM) regime, we employed the cell model \cite{Wu2021,Wu2023}, which enforces the ideal-gas limit \(C^{ion}_{V} \rightarrow \frac{3}{2}k_{\text{B}}\ \text{per atom}\) as \(T \rightarrow \infty\). This free-energy framework is given by the following expressions:
\begin{equation}
\begin{aligned}
F_{\text{vib,liquid}}(\rho, T)
={}& F_{\text{vib,solid}}(\rho, T)
- k_{\text{B}}T \ln \Biggl[
\operatorname{erf}\!\left( \frac{T^*}{T} \right) \\
&\qquad\qquad
-\frac{2}{\sqrt{\pi}} \sqrt{\frac{T^*}{T}}
\exp\!\left(-\frac{T^*}{T}\right)
\Biggr],
\end{aligned}
\end{equation}
where \(k_{B}T^{*}\) is the $\rho$-dependent energy defined in the cell-model  \cite{Wu2021,Wu2023}.
\[ k_{\text{B}}T^*(\rho) = \frac{M k_{\text{B}}^2 \left[ \Theta_D(\rho) R(\rho) \right]^2}{2\hbar^2},\]
$\Theta_D(\rho)$ is $\rho$-dependent Debye temperature derived from DFT $\rho$-dependent elastic constants (Fig.~\ref{fig:DFT-Debye}) and
\[ R(\rho) = \left( \frac{3M}{4\pi\rho} \right)^{\frac{1}{3}},\] where $M$ is the mass per atom (i.e., the atomic mass).

For the internal energy, we adopt the following form so that the ideal-gas limit is recovered as \(T\rightarrow\infty\):
\begin{equation}
\begin{aligned}
U_{\text{vib,liquid}}(\rho, T)
={}& U_{\text{vib,solid}}(\rho, T)
\left[ \frac{1}{2}+\frac{1}{2}\exp\!\left(-\sqrt{\frac{T}{T^*}}\right) \right].
\end{aligned}
\end{equation}

This construction provides a continuous description of the liquid vibrational internal energy and free energy from low-temperature liquid states into the warm-dense-matter regime.

%===================================================================================
\section{Uncertainty sources in DFT-based EOS development}\label{sec:DFTuncertainty}
%===================================================================================

Uncertainty quantification (UQ) within DFT is crucial for evaluating the reliability and accuracy of EOS data obtained from simulations. 
Sources of uncertainty include the choice of exchange–correlation functional, the treatment of pseudopotentials and relativistic effects, numerical approximations used to solve the Kohn–Sham equations, supercell finite-size effects, and the specific convergence thresholds adopted. 
Extensive research has focused on characterizing and mitigating these sources of error \cite{Nagai2020,Wang2020Bayesian,Wang2020DFT,Gygi2023}.
In UEOS, we differentiate between two main categories of uncertainty that enter the GP training likelihood:
(i) \emph{model discrepancy} between a DFT-based description and experimental reference data (denoted as ``Exp.--Theory''), and 
(ii) \emph{internal theoretical/numerical uncertainty} arising from the DFT workflow itself (denoted as ``Theory'').
For each free-energy contribution $g\in\{F_{\mathrm{cold}},F_{\mathrm{elec}},F_{\mathrm{vib}}\}$, we specify a relative uncertainty $r_g$, which we then use to define an output-noise model for each training point,
\begin{equation}
\sigma_{y,i} = r_g\,|y_i|.
\label{eq:rel_uncertainty}
\end{equation}
When several (assumed independent) sources of uncertainty are present, we add the corresponding relative uncertainties in quadrature, for example,
$r_{g,\mathrm{tot}}=\sqrt{r_{g,\mathrm{Exp-Theory}}^2+r_{g,\mathrm{Theory}}^2}$.
This global relative-uncertainty model serves as a convenient baseline; UEOS can also accommodate pointwise (heteroscedastic) uncertainty specifications when such information is available.

Benchmarking against experimental data and/or theoretical approaches provides an empirical foundation for quantifying model discrepancy. 
For the cold curve, for example, we compare DFT isotherms with DAC and toroidal-DAC measurements (Fig.~\ref{fig:DFT-DAC}); the resulting percentage differences justify the conservative choice of $r_{F_{\mathrm{cold}},\mathrm{Exp-Theory}}=5\%$ adopted in this work. 
By contrast, direct experimental constraints on the thermal free-energy terms $F_{\mathrm{elec}}$ and $F_{\mathrm{vib}}$ are relatively limited. 
Consequently, the uncertainty estimates listed in Table~\ref{tab:Au_DFTuncertainty} are based on (i) the variation introduced by different electronic-structure treatments (exchange–correlation functional and relativistic scheme), (ii) the dependence on ionic configurations in finite-temperature electronic calculations, and (iii) the level of agreement between computed and experimentally inferred Debye temperatures (Fig.~\ref{fig:DFT-Debye}) and ambient-condition specific heat.

Internal theoretical uncertainties are estimated from convergence and sensitivity studies within our DFT workflows. 
For the vibrational contribution, the dominant terms include supercell-size effects (we compared 64, 108, and 256 atom cells), configuration sampling in BOMD-based liquid references, and higher-order terms beyond the quasi-harmonic approximation in SCP calculations. 
For the electron-thermal term, variations across ionic snapshots provide an estimate of uncertainty associated with finite-$T$ electronic occupation at fixed $(\rho,T)$. For bookkeeping, Table~\ref{tab:Au_DFTuncertainty} separates the vibrational contribution into a quasi-harmonic part and an anharmonic correction.

\begin{table}[t]
\centering
\caption{Relative uncertainties assigned to the DFT-based free-energy training data employed in the construction of the Au UEOS table. ``Exp.--Theory'' denotes the theory--experiment (model) discrepancy with respect to available experimental reference constraints. ``Theory'' denotes internal numerical and theoretical uncertainty associated with convergence behavior, sampling procedures, and the choice of DFT exchange--correlation functional. ``Combined'' corresponds to the quadrature sum of these contributions and is used to define $\sigma_{y,i}$ in Eq.~\eqref{eq:rel_uncertainty}.}
\label{tab:Au_DFTuncertainty}
\begin{tabular}{lcccc}
\toprule
Source of uncertainty & $F_{\mathrm{cold}}$ & $F_{\mathrm{elec}}$ & $F_{\mathrm{vib,QHA}}$ & $\Delta F_{\mathrm{vib,ah}}$ \\
\midrule
Exp.--Theory & 5\% & 1\% & 3\% & -- \\
Theory       & 1\% & 1\% & 2\% & 4\% \\
Combined     & 5.1\% & 1.4\% & 3.6\% & 4\% \\
\bottomrule
\end{tabular}
\end{table}

Table~\ref{tab:Au_DFTuncertainty} provides a concise overview of the uncertainty model adopted in this work. 
In practice, the associated uncertainties are not strictly uniform over the $(\rho,T)$ state space, and correlations between neighboring state points may arise (for example, due to systematic biases from specific density functionals or shared numerical parameters). 
A rigorous treatment of heteroscedasticity and correlation remains a key direction for future work, for example through the use of nonstationary covariance kernels \cite{PaciorekSchervish2006}, sparse heteroscedastic GP approaches \cite{Almosallam2016GPz}, or hierarchical and deep GP architectures \cite{GoncalvesWellmann2025}.
%===================================================================================
\section{Uncertainty-aware EOS for gold}
\label{sec:UEOS}

\begin{figure}[htbp]
    \centering
          \includegraphics[width=\columnwidth, trim=0cm 8cm 0cm 4cm,clip]{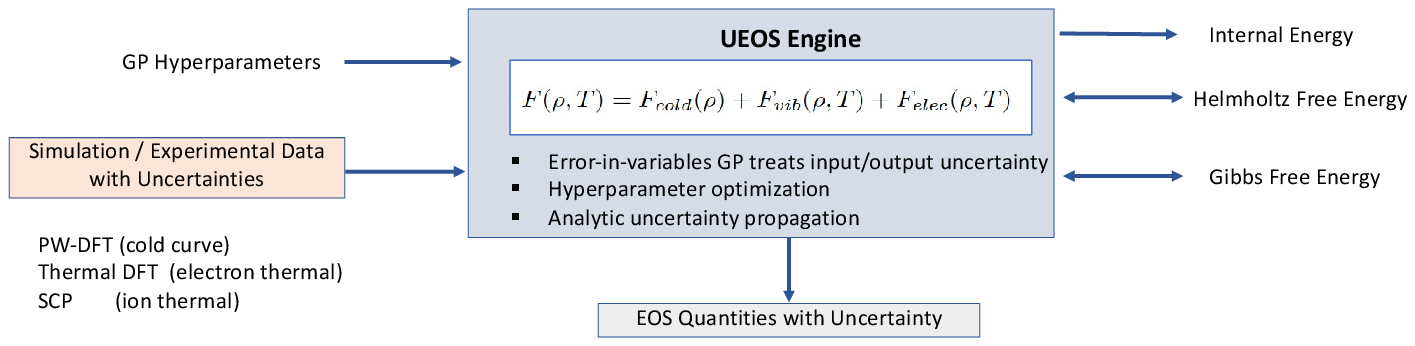}

    \caption{Schematic flowchart illustrating the general UEOS workflow. Simulation- or experiment-derived data, together with their associated uncertainties, specify a GP regression problem and an accompanying optimization procedure for hyperparameter estimation. The resulting trained GP model(s) are subsequently employed in uncertainty-propagation (UP) operations---such as differentiation and algebraic manipulation---to construct free-energy surfaces and thermodynamic response functions, along with their corresponding predictive uncertainty estimates.}
    \label{fig:workflow}
\end{figure}

The EOS table is constructed by evaluating the free energy on a prescribed $(\rho,T)$ grid using UEOS (Fig.~\ref{fig:workflow}). 
At each state point, thermodynamic properties are obtained from derivatives of the Helmholtz free energy. For example, the thermal pressure is
\begin{equation}
P(\rho,T)=\frac{\rho^{2}}{M}\left(\frac{\partial F(\rho,T)}{\partial\rho}\right)_{T}.
\label{eq:P_from_F}
\end{equation}

Because the GP posterior is differentiable, $\left(\partial F/\partial\rho\right)_T$ can be evaluated analytically, and its uncertainty propagated to $P$ (Fig.~\ref{fig:Au_cold_EOS}). 

The Gibbs free energy follows from
\begin{equation}
\begin{aligned}
G(\rho,T)
={}&F(\rho,T)+P(\rho,T)\frac{M}{\rho} \\
={}&F(\rho,T)+\rho\left(\frac{\partial F(\rho,T)}{\partial\rho}\right)_{T}.
\end{aligned}
\label{eq:G_from_F}
\end{equation}
Equations~\eqref{eq:P_from_F}--\eqref{eq:G_from_F} illustrate how UEOS provides a unified route to mean predictions and uncertainty estimates for derived EOS quantities starting from a GP representation of $F(\rho,T)$.

\subsection{Helmholtz free energy with predicted uncertainties}

\begin{figure}[htbp]
    \centering
         \includegraphics[scale=0.40,trim=0cm 4cm 0cm 0cm,clip]{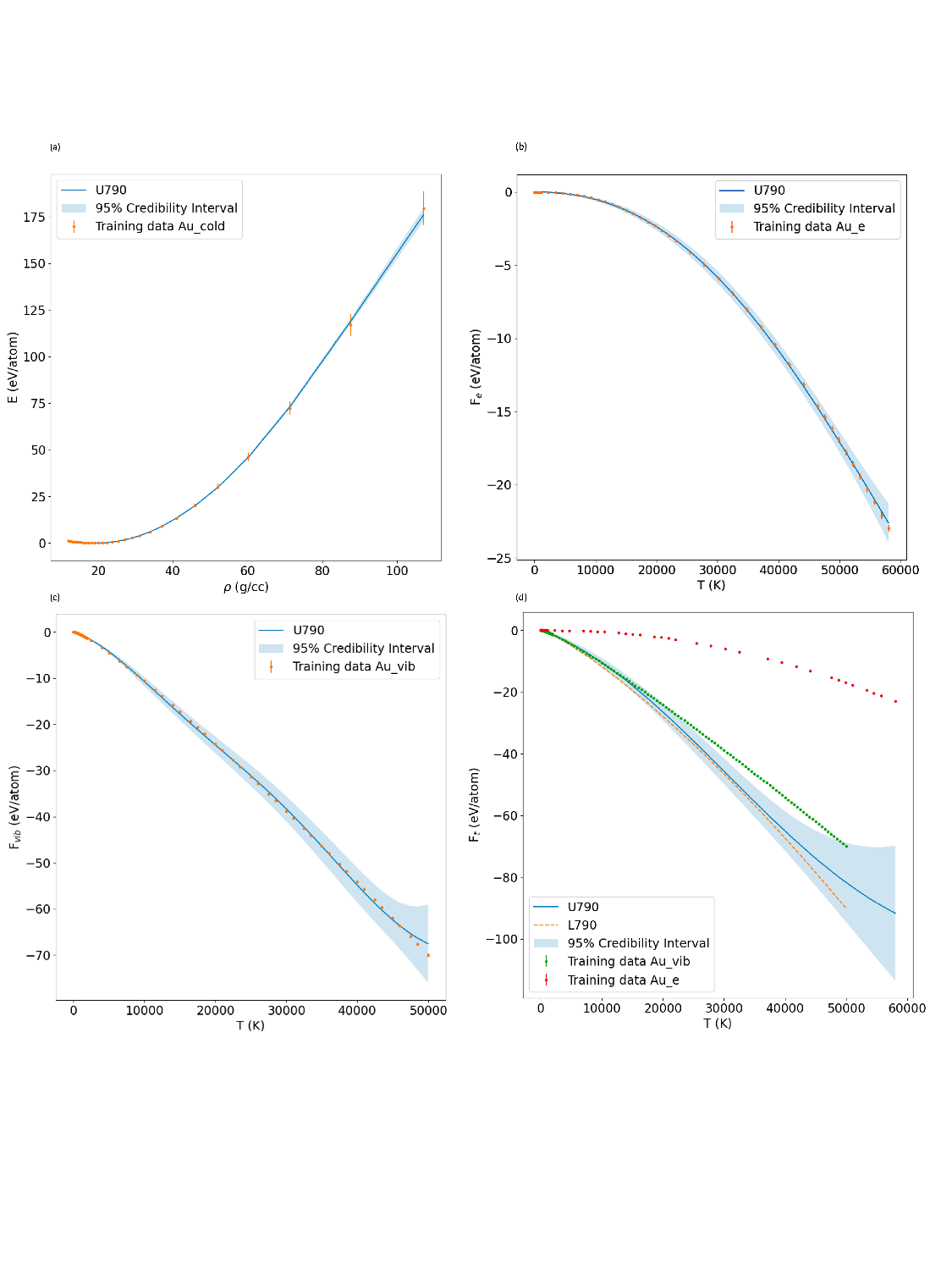}

    \caption{Construction of the total Helmholtz free energy with associated uncertainty using the UEOS framework. Separate EIV--GP models are trained for each free-energy contribution: (a) cold, (b) electron-thermal, and (c) ion-thermal, with the corresponding uncertainty model summarized in Table~\ref{tab:Au_DFTuncertainty}. These contributions are subsequently combined to produce (d) the total free energy, including the propagated predictive uncertainty. The labels ``Au\_cold'', ``Au\_e'', and ``Au\_vib'' denote, respectively, the free-energy datasets for the cold, electron-thermal, and ion-thermal contributions.}
    \label{fig:UEOS-Ft}
\end{figure}

Figure~\ref{fig:UEOS-Ft} illustrates the construction of the Helmholtz free energy $F(\rho = 2\rho_0, T)$ over the temperature range $0 \le T \le 5 \times 10^{4}$~K by combining GP models for the cold, electron-thermal, and ion-thermal contributions. 
The shaded band denotes the posterior credible interval implied by the uncertainty model described in Sec.~\ref{sec:DFTuncertainty}. 
As expected, the predictive uncertainty is smallest where the training data are dense and grows in sparsely sampled or extrapolative regions. 
As discussed in Sec.~\ref{sec:GP}, physics-informed mean functions---or, equivalently, GP models for residuals relative to a baseline EOS---provide a practical route for reducing extrapolative variance when needed.

\begin{figure}[htbp]
    \centering
        \includegraphics[width=\columnwidth, trim=0cm 1cm 0cm 0cm,clip]{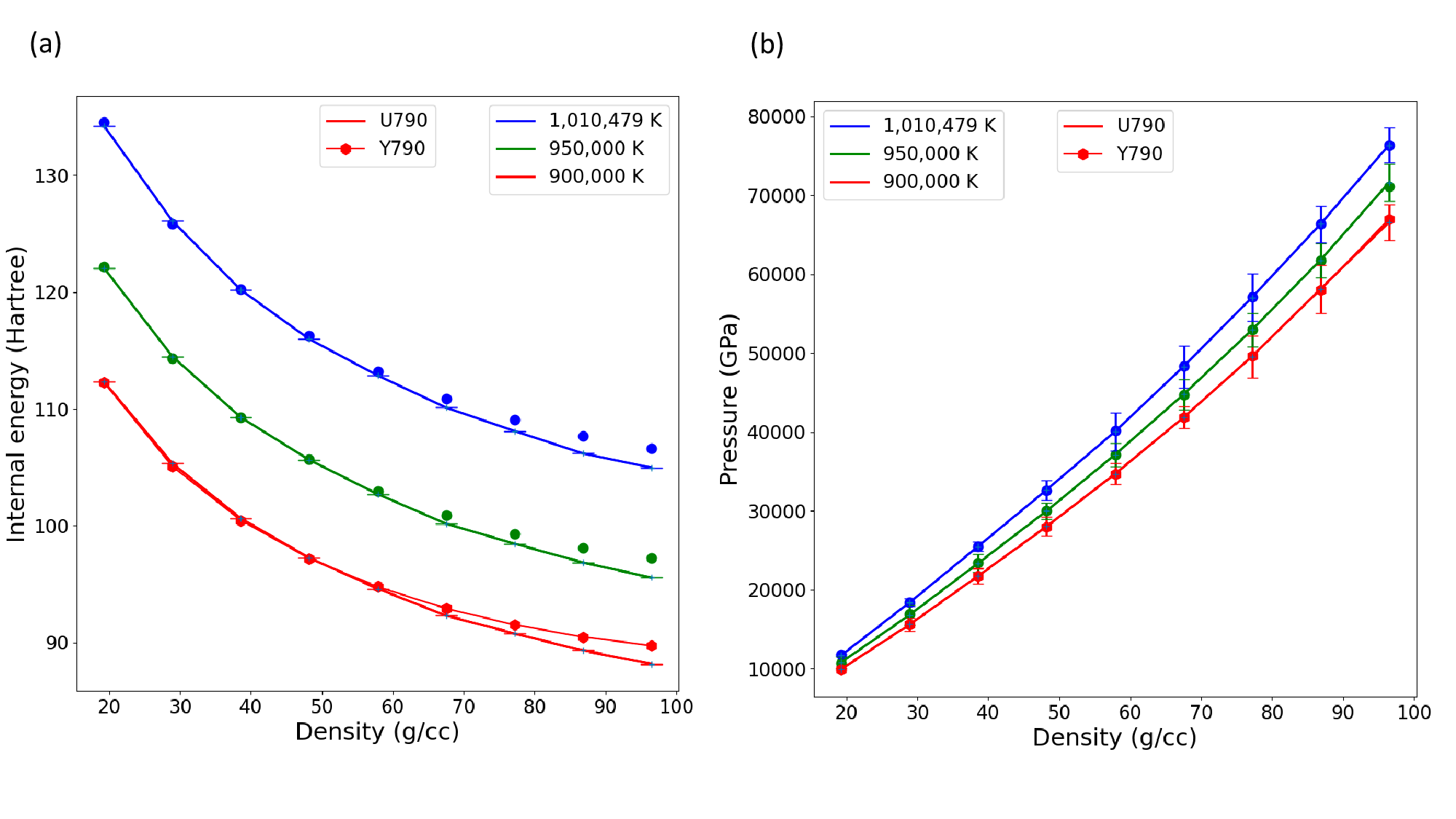}

    \caption{(a) Comparison of the internal energy of U790 and Y790 for $19 < \rho < 100$ g/cc at $T = 9\times10^{5}$ K, $9.5\times10^{5}$ K, and $1.010479\times10^{6}$ K. (b) Comparison of the total thermal pressure of U790 and Y790 at the same temperatures.}
    \label{fig:U790-Y790_UP}
\end{figure}

\subsection{Comparisons with existing LLNL EOS tables}

\begin{figure}[htbp]
    \centering
        \includegraphics[width=\columnwidth, trim=0cm 0.5cm 0cm 2cm,clip]{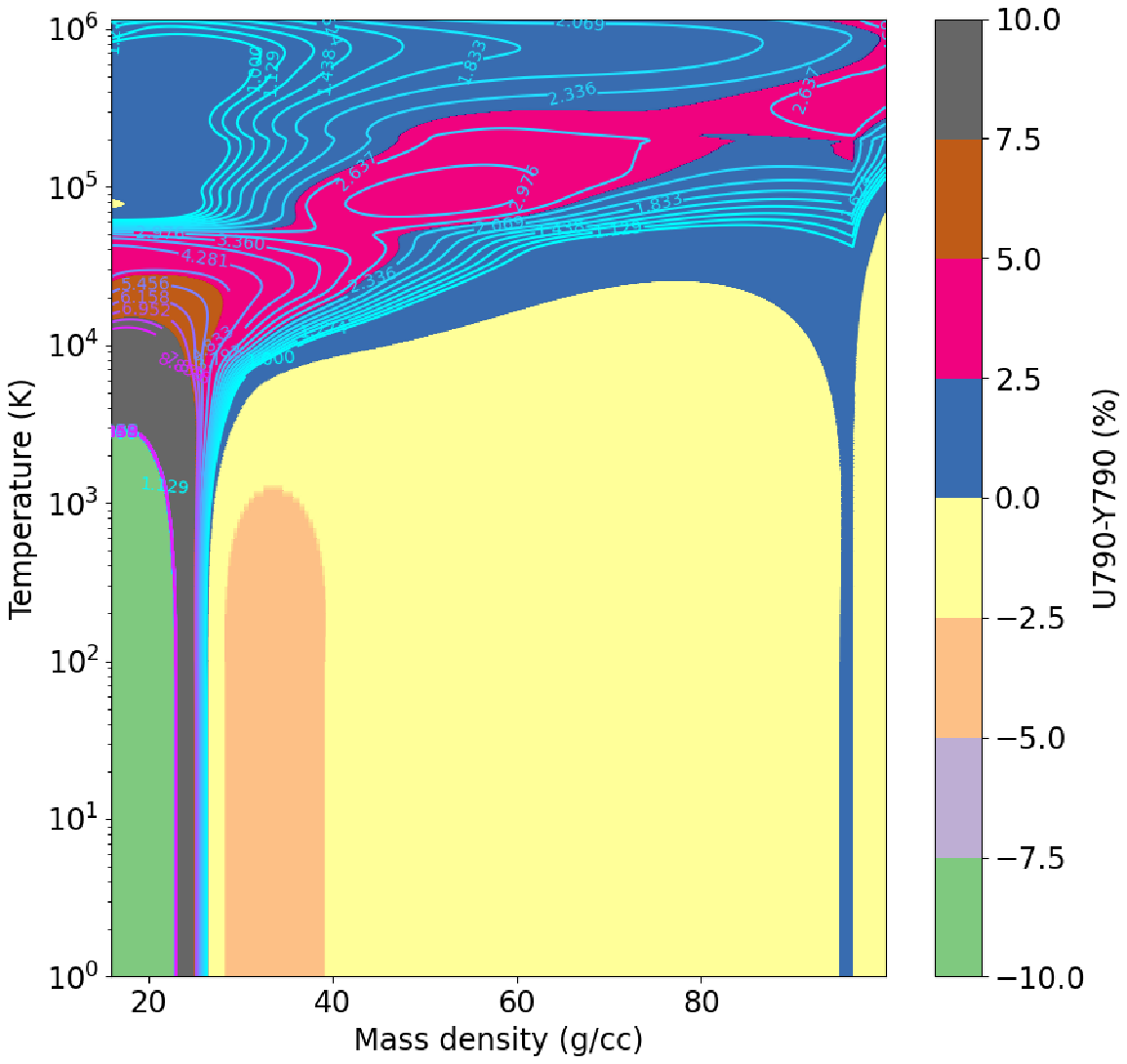}

    \caption{Helmholtz free-energy comparison between U790 and Y790 for $16 < \rho < 100$ g/cc and $0 \le T \le 10^{6}$ K. The two remain consistent to within $\lesssim 10\%$ over most of the plotted domain, with systematic deviations that arise from differences in the underlying physical models and calibration choices.}
    \label{fig:U790-Y790_Ft_grid}
\end{figure}

To assess the performance of U790, we compare it against the established LLNL Au EOS tables L790 and Y790. 
The L790 table employs a Thomas--Fermi model to describe the electron-thermal contribution, whereas the Y790 table is based on the Purgatorio model; these distinct treatments of electronic thermal effects give rise to systematic differences at elevated temperatures. 
The thermodynamic-grid comparisons in Figs.~\ref{fig:U790-Y790_UP} and~\ref{fig:U790-Y790_Ft_grid} focus on Y790, while the shock-Hugoniot comparisons below include both L790 and Y790. Together, these results show that U790 generally agrees with the existing LLNL tables to within $\mathcal{O}(10\%)$ over a substantial fraction of their overlapping thermodynamic domain, while also revealing structured regions in which the deviations are significantly larger.

The non-random spatial patterns visible in Fig.~\ref{fig:U790-Y790_Ft_grid} constitute a useful diagnostic: they help identify regimes in which the dominant source of uncertainty is likely attributable to \emph{model-form} discrepancy (e.g., deficiencies or inconsistencies in the representation of electron–thermal physics), rather than arising solely from parametric variability.  
Within the UEOS framework, such maps can be utilized to (i) inform the targeted acquisition of additional training data (through focused DFT calculations or purpose-designed experiments) and (ii) provide quantitative justification for introducing more physics-informed prior means in selected regions of the (\(\rho,T\)) state space.

\subsection{Shock Hugoniot comparison and quantitative consistency checks}
\begin{figure}[htbp]
    \centering
        \includegraphics[width=\columnwidth, trim=0cm 0cm 0cm 0cm,clip]{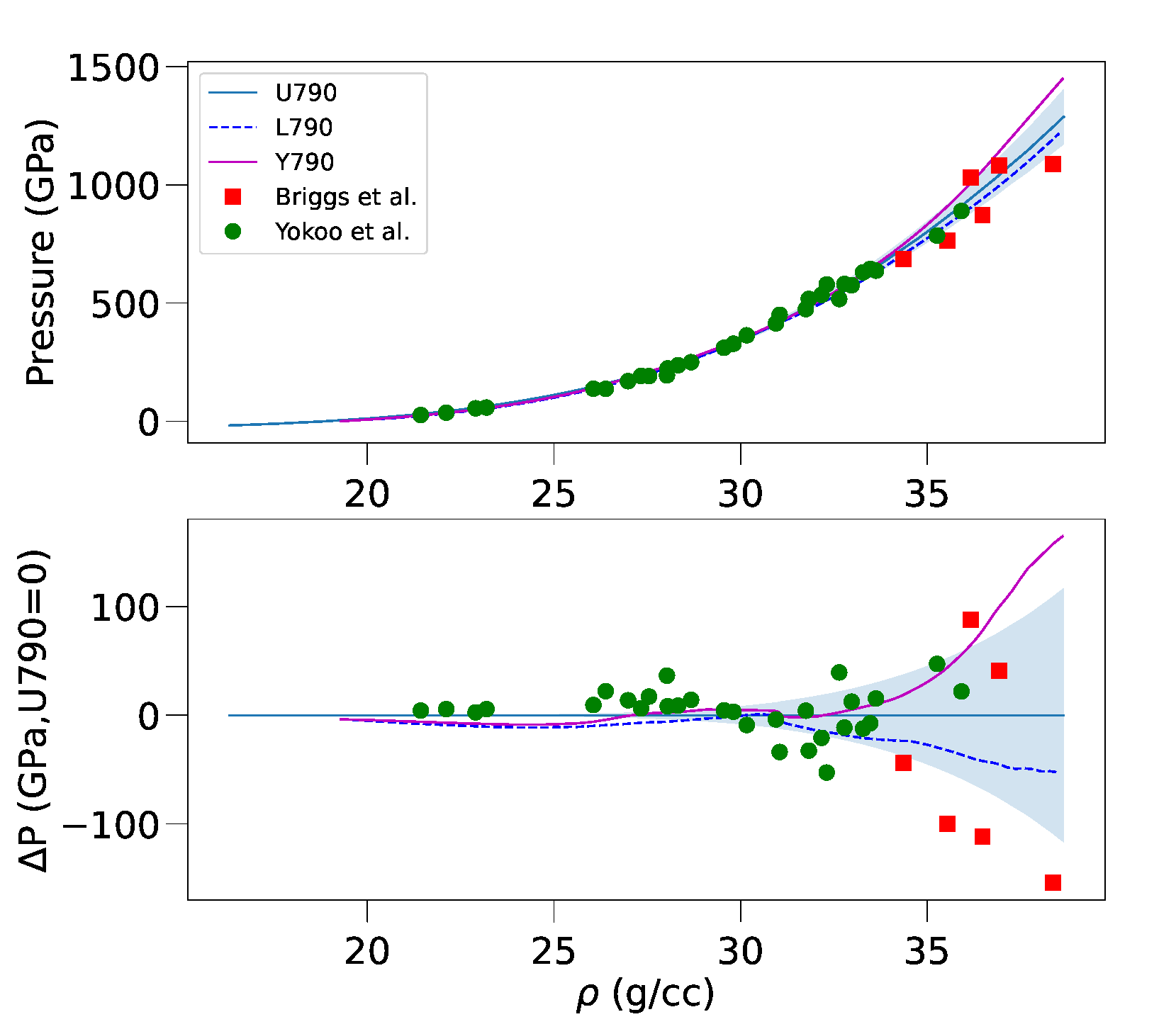}

    \caption{Shock Hugoniot comparison (pressure versus density) from initial density $\rho_0=19.28$~g/cc. We compare the present UEOS table (U790) with L790, Y790, and experimental measurements \cite{Yokoo2008,Briggs2019}. The shaded region indicates the U790 predictive credible interval propagated from the free-energy model.}
    \label{fig:Au_Hugoniot}
\end{figure}

\begin{figure}[htbp]
    \centering
          \includegraphics[width=\columnwidth, trim=0cm 0cm 0cm 0cm,clip]{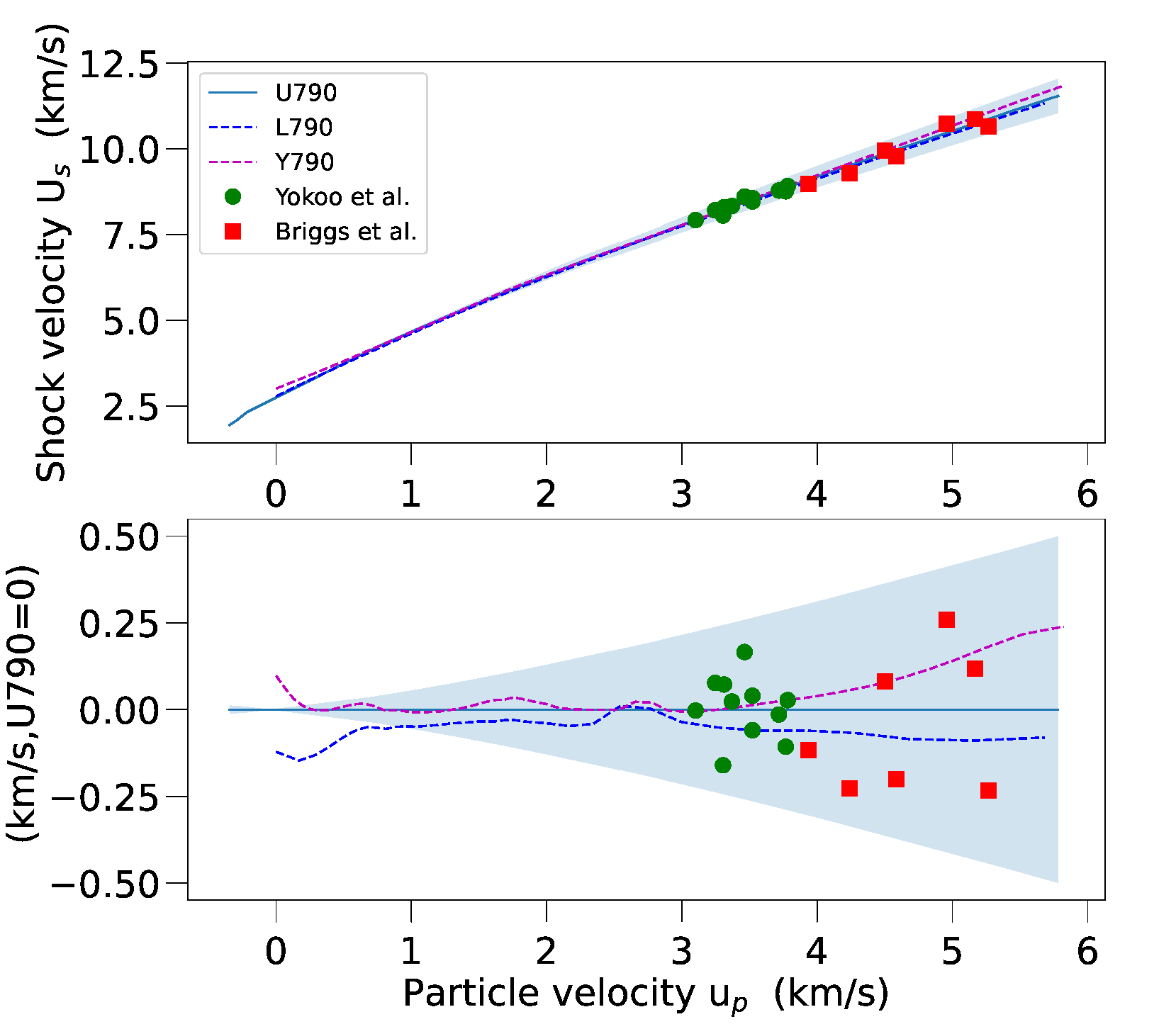}

    \caption{Shock Hugoniot $U_s$--$u_p$ relation from initial density $\rho_0=19.28$~g/cc comparing U790 with L790, Y790, and experiments \cite{Yokoo2008,Briggs2019}. The shaded region indicates the U790 predictive credible interval.}
    \label{fig:Au_Hugoniot_Usup}
\end{figure}

A primary way to verify an EOS table is by comparing it with the shock Hugoniot, which is obtained by solving the Rankine–Hugoniot equation using the tabulated values of pressure, density, and internal energy. Figures~\ref{fig:Au_Hugoniot}--\ref{fig:Au_Hugoniot_Usup} demonstrate that U790 accurately reproduces the established Hugoniot response of Au and shows consistency with both existing LLNL EOS tables and experimental data \cite{Yokoo2008,Briggs2019}.


When experimental uncertainty intervals are available, UEOS can incorporate them explicitly (through $\Sigma_\mathbf{X}$ and $\sigma_y$) within an EIV regression framework or in subsequent consistency and validation analyses. A practical quantitative diagnostic is to construct normalized residuals,
\begin{equation}
r_i=\frac{P_{\mathrm{exp}}(\rho_i)-P_{\mathrm{U790}}(\rho_i)}{\sqrt{\sigma_{\mathrm{exp}}(\rho_i)^2+\sigma_{\mathrm{U790}}(\rho_i)^2}},
\end{equation}
which are expected to be approximately standard normally distributed if the uncertainty estimates are well calibrated and mutually independent. Additional, complementary assessments include coverage analyses (i.e., evaluating the fraction of experimental observations that lie within a nominal $95\%$ credible interval) and distributional goodness-of-fit tests, such as Kolmogorov--Smirnov statistics applied to the set $\{r_i\}$. Collectively, these diagnostics provide a rigorous quantitative linkage between the uncertainty-aware EOS representation and model validation against experimental measurements.

%===================================================================================
\section{Conclusions}
\label{sec:conclusions}

We present the UEOS framework, an uncertainty-aware EOS workflow that combines GP regression with an EIV likelihood, allowing uncertainty in the training data to be propagated through to smooth, differentiable thermodynamic potentials and the resulting EOS properties. 
The framework is designed to handle uncertainty in both the outputs (e.g., free energies and pressures) and the inputs (e.g., density and temperature) within a computationally feasible probabilistic formulation, and it reduces to standard GP regression in the limit where input uncertainty is negligible.

We demonstrated the methodology by generating an uncertainty-aware EOS table for gold (U790). This table is constructed from first-principles DFT free-energy contributions, augmented with an explicit statistical representation of the uncertainties in the training data (Table~\ref{tab:Au_DFTuncertainty}), on a $(\rho,T)$ grid comprising 121 density points and 2000 temperature points, with $\rho$ sampled linearly and $T$ sampled on a logarithmic scale. 
The resulting EOS table delivers mean predictions together with state-dependent uncertainty estimates across a broad range of densities (reaching compressions up to $\rho = 100$~g/cc) and temperatures (up to approximately 300~eV). 
Benchmarking against established LLNL EOS tables and shock-compression experiments shows that U790 faithfully reproduces the known Hugoniot behavior while supplying realistic uncertainty bands that are appropriate for subsequent uncertainty propagation and model validation.

Exact GP training scales as $\mathcal{O}(N^3)$ with respect to the number of training points $N$, and requires $\mathcal{O}(N^2)$ memory for storage and factorization of the covariance matrix. 
In the Au case study considered here, the individual component datasets are sufficiently small that exact GP training remains computationally feasible. 
However, when the methodology is applied to much larger databases involving multiple materials or fidelities, it becomes highly advantageous to rely on approximate inference strategies. UEOS is compatible with standard sparse and inducing-point GP approximations \cite{Snelson2006,Titsias2009,QuinoneroRasmussen2005}, as well as with heteroscedastic sparse GP models such as GPz \cite{Almosallam2016GPz}, which substantially decrease computational costs while maintaining well-calibrated uncertainty estimates.

A primary motivation for developing uncertainty-aware EOS tables is their use in downstream simulations and in planning experiments that rely on hydrodynamic calculations. Since UEOS provides a differentiable representation of $F(\rho,T)$ together with a predictive probability distribution, simulation frameworks can (i) adopt the mean EOS as a nominal input for deterministic runs, (ii) sample EOS realizations from the predictive distribution to carry out Monte Carlo propagation of EOS uncertainty into derived observables, or (iii) leverage local uncertainty estimates to direct new measurements or simulations toward regions of state space where uncertainty most strongly impacts quantities of interest. Future developments will target (1) more explicit enforcement of thermodynamic consistency during GP training \cite{Sharma2024ThermoEOS}, (2) improvement of the uncertainty model beyond a single global relative-noise assumption to represent heteroscedasticity and spatial or parametric correlations \cite{PaciorekSchervish2006}, and (3) generalizing UEOS to support joint training on heterogeneous datasets that integrate experimental data and simulation results.

%=================================================================================== 
\section*{Acknowledgments}
%===================================================================================

We express our gratitude to Drs. Suzanne Ali, Eric Herbold, John Pask, Per Söderlind, and Phil Sterne for their insightful discussions and feedback on this manuscript. This research was conducted under the auspices of the U.S. Department of Energy by Lawrence Livermore National Laboratory pursuant to Contract No. DE-AC52-07NA27344.

\appendix

\section{Error-in-variables Gaussian processes}
\label{app:eivgp}

This appendix summarizes the EIV--GP regression used in UEOS. The formulation is standard in the GP literature \cite{Girard2003,Girard2005,McHutchon2011,Johnson2020,Rasmussen2005} and is included here to make clear (i) how input uncertainty enters the likelihood and (ii) how analytic kernel derivatives are used for thermodynamic derivatives and uncertainty propagation.

\subsection{Effective likelihood with uncertain inputs}

Let the training set contain $N$ observations $\{(\mathbf{X}_i,y_i)\}_{i=1}^{N}$ with $\mathbf{X}_i\in\mathbb{R}^{D}$. We assume
\begin{align}
y_i &= f(\tilde{\mathbf{X}}_i)+\epsilon_{y,i}, \qquad \epsilon_{y,i}\sim\mathcal{N}(0,\sigma_{y,i}^{2}), \\
\mathbf{X}_i &= \tilde{\mathbf{X}}_i+\boldsymbol{\epsilon}_{\mathbf{X},i}, \qquad \boldsymbol{\epsilon}_{\mathbf{X},i}\sim\mathcal{N}(\mathbf{0},\Sigma_{\mathbf{X},i}),
\end{align}
where $\tilde{\mathbf{X}}_i$ are latent inputs and $\Sigma_{\mathbf{X},i}$ is an input-noise covariance (often taken diagonal).

Using a first-order Taylor expansion,
\begin{equation}
f(\tilde{\mathbf{X}}_i+\boldsymbol{\epsilon}_{\mathbf{X},i})\approx f(\tilde{\mathbf{X}}_i)+\nabla f(\tilde{\mathbf{X}}_i)^{\mathsf{T}}\boldsymbol{\epsilon}_{\mathbf{X},i},
\end{equation}
the input uncertainty contributes an additive variance term to the effective observation (output) noise,
\begin{equation}
\sigma_{\mathrm{eff},i}^{2}\approx \sigma_{y,i}^{2}+\nabla f(\tilde{\mathbf{X}}_i)^{\mathsf{T}}\Sigma_{\mathbf{X},i}\nabla f(\tilde{\mathbf{X}}_i).
\label{eq:sigma_eff}
\end{equation}
Collecting these into $\Sigma_{\mathrm{eff}}=\mathrm{diag}(\sigma_{\mathrm{eff},1}^{2},\ldots,\sigma_{\mathrm{eff},N}^{2})$, define
\begin{equation}
C_\theta = K(\mathbf{X},\mathbf{X};\theta)+\Sigma_{\mathrm{eff}}.
\end{equation}
The GP predictive mean and variance at a test input $\mathbf{X}_*$ are then
\begin{align}
\mu_* &= \mathbf{k}_*^{\mathsf{T}}C_\theta^{-1}\mathbf{y},\\
\sigma_*^{2} &= k(\mathbf{X}_*,\mathbf{X}_*)-\mathbf{k}_*^{\mathsf{T}}C_\theta^{-1}\mathbf{k}_*,
\end{align}
where $[K(\mathbf{X},\mathbf{X};\theta)]_{ij}=k(\mathbf{X}_i,\mathbf{X}_j)$, $\mathbf{k}_*=[k(\mathbf{X}_1,\mathbf{X}_*),\ldots,k(\mathbf{X}_N,\mathbf{X}_*)]^{\mathsf{T}}$, and $\mathbf{y}=[y_1,\ldots,y_N]^{\mathsf{T}}$.

\subsection{Marginal likelihood and iterative training}

With $\Sigma_{\mathrm{eff}}$ held fixed, the (log) marginal likelihood is
\begin{equation}
\log p(\mathbf{y}\mid \mathbf{X},\theta)
=-\frac12 \left[
\mathbf{y}^{\mathsf{T}}C_\theta^{-1}\mathbf{y}
+\log\lvert C_\theta\rvert
+N\log(2\pi)\right].
\label{eq:log_marginal}
\end{equation}
In EIV regression, the matrix $\Sigma_{\mathrm{eff}}$ depends on the gradient $\nabla f$ via Eq.~\eqref{eq:sigma_eff}, which renders the likelihood function implicit. A widely used and effective strategy is to employ an outer fixed-point iteration \cite{McHutchon2011,Johnson2020}:
\begin{enumerate}
\item Initialize the kernel hyperparameters $\theta^{(0)}$ and set $\Sigma_{\mathrm{eff}}^{(0)}=\mathrm{diag}(\sigma_{y,1}^{2},\ldots,\sigma_{y,N}^{2})$.
\item For iterations $t=0,1,2,\ldots$:
\begin{enumerate}
\item Optimize $\theta^{(t+1)}$ by maximizing $\log p(\mathbf{y}\mid \mathbf{X},\theta)$ in Eq.~\eqref{eq:log_marginal}, holding $\Sigma_{\mathrm{eff}}^{(t)}$ fixed.
\item Compute the posterior mean gradient $\nabla \mu(\mathbf{X}_i)$ at the training inputs.
\item Update $\Sigma_{\mathrm{eff}}^{(t+1)}$ using Eq.~\eqref{eq:sigma_eff}, with the approximation $\nabla f(\tilde{\mathbf{X}}_i)\approx \nabla \mu(\mathbf{X}_i)$.
\end{enumerate}
\item Terminate the iteration once the changes in $\theta$ and $\Sigma_{\mathrm{eff}}$ fall below a prescribed tolerance.
\end{enumerate}
This iterative scheme makes explicit the treatment of input uncertainties: the covariances $\Sigma_{\mathbf{X},i}$ are not optimized as free parameters in this work but are instead prescribed by the user based on experimental or simulation-derived uncertainty estimates, whereas $\theta$ denotes the GP hyperparameters that are inferred from the data.

\subsection{Squared-exponential kernel derivatives}

For the squared-exponential (SE) kernel,
\begin{equation}
k(\mathbf{X},\mathbf{X}')=\sigma_f^{2}\exp\!\left[-\frac12(\mathbf{X}-\mathbf{X}')^{\mathsf{T}}\Lambda^{-1}(\mathbf{X}-\mathbf{X}')\right],
\end{equation}
with $\Lambda=\mathrm{diag}(\lambda_1^{2},\ldots,\lambda_D^{2})$ and $\Delta\mathbf{X}=\mathbf{X}-\mathbf{X}'$, the gradient with respect to $\mathbf{X}$ is
\begin{equation}
\nabla_{\mathbf{X}}k(\mathbf{X},\mathbf{X}')=-\,k(\mathbf{X},\mathbf{X}')\,\Lambda^{-1}\Delta\mathbf{X},
\end{equation}
and $\nabla_{\mathbf{X}'}k(\mathbf{X},\mathbf{X}')=k(\mathbf{X},\mathbf{X}')\,\Lambda^{-1}\Delta\mathbf{X}$.
The mixed second derivative ($D\times D$) is
\begin{equation}
\nabla_{\mathbf{X}}\nabla_{\mathbf{X}'}^{\mathsf{T}}k(\mathbf{X},\mathbf{X}')
= k(\mathbf{X},\mathbf{X}')
\left[
\Lambda^{-1}-\Lambda^{-1}\Delta\mathbf{X}\,\Delta\mathbf{X}^{\mathsf{T}}\Lambda^{-1}
\right].
\end{equation}
These expressions are used to evaluate thermodynamic derivatives (e.g., $\partial F/\partial\rho$ and $\partial F/\partial T$) and to propagate uncertainty from $F(\rho,T)$ to derived quantities such as pressure.

\subsection{Mat\'ern kernel derivatives}
\label{app:matern}

In some EOS applications it can be desirable to relax the infinite smoothness implied by the squared-exponential kernel and instead control the differentiability of the GP through a Mat\'ern covariance. The Mat\'ern family is parameterized by a smoothness parameter $\nu>0$ and reduces to increasingly smooth processes as $\nu$ increases \cite{Rasmussen2005}. Derivatives of the GP exist in the mean-square sense only up to order $m<\nu$; thus, for first derivatives (e.g., pressure from $\partial F/\partial\rho$) one typically chooses $\nu>1$ (e.g., $\nu=3/2$), and for second derivatives (e.g., heat capacities) one typically chooses $\nu>2$ (e.g., $\nu=5/2$).

As in the SE case, define
\begin{equation}
\begin{aligned}
r^{2}(\mathbf{X},\mathbf{X}')
={}&(\mathbf{X}-\mathbf{X}')^{\mathsf{T}}\Lambda^{-1}(\mathbf{X}-\mathbf{X}'), \\
\Lambda
={}&\mathrm{diag}(\lambda_1^2,\ldots,\lambda_D^2).
\end{aligned}
\end{equation}
With $\alpha=\sqrt{2\nu}\,r$ and $\Delta\mathbf{X}=\mathbf{X}-\mathbf{X}'$, the Mat\'ern kernel can be written
\begin{equation}
k_{\nu}(\mathbf{X},\mathbf{X}')=\sigma_f^2\,
\frac{2^{1-\nu}}{\Gamma(\nu)}\,\alpha^{\nu}K_{\nu}(\alpha),
\label{eq:matern_kernel}
\end{equation}
where $K_{\nu}$ is the modified Bessel function of the second kind. Using the identity
$\frac{\mathrm{d}}{\mathrm{d}\alpha}\big[\alpha^{\nu}K_{\nu}(\alpha)\big]=-\alpha^{\nu}K_{\nu-1}(\alpha)$,
the radial derivatives are
\begin{align}
k_{\nu}'(r)\equiv \frac{\mathrm{d}k_{\nu}}{\mathrm{d}r}
&= -\sigma_f^2\,\frac{2^{1-\nu}}{\Gamma(\nu)}\,\sqrt{2\nu}\,\alpha^{\nu}K_{\nu-1}(\alpha),
\label{eq:matern_kprime}\\
k_{\nu}''(r)\equiv \frac{\mathrm{d}^2k_{\nu}}{\mathrm{d}r^2}
&= \sigma_f^2\,\frac{2^{1-\nu}}{\Gamma(\nu)}\,(2\nu) \\
&\quad\times\left[\alpha^{\nu}K_{\nu-2}(\alpha)-\alpha^{\nu-1}K_{\nu-1}(\alpha)\right],
\label{eq:matern_ksecond}
\end{align}
where Eq.~\eqref{eq:matern_ksecond} follows from differentiating Eq.~\eqref{eq:matern_kprime} and using the Bessel-function identity
$\frac{\mathrm{d}}{\mathrm{d}\alpha}\!\left[\alpha^{\nu}K_{\nu-1}(\alpha)\right]=\alpha^{\nu-1}\!\left[\alpha K_{\nu-2}(\alpha)-K_{\nu-1}(\alpha)\right]$.

The gradient with respect to $\mathbf{X}$ is obtained by the chain rule,
\begin{align}
\nabla_{\mathbf{X}}k_{\nu}(\mathbf{X},\mathbf{X}')
= k_{\nu}'(r)\,\frac{\Lambda^{-1}(\mathbf{X}-\mathbf{X}')}{r},
\\
\nabla_{\mathbf{X}'}k_{\nu}(\mathbf{X},\mathbf{X}')
= -\,k_{\nu}'(r)\,\frac{\Lambda^{-1}(\mathbf{X}-\mathbf{X}')}{r}.
\label{eq:matern_grad}
\end{align}
For covariance calculations involving derivatives, the mixed second derivative ($D\times D$) can be written compactly in terms of $k_\nu'(r)$ and $k_\nu''(r)$:
\begin{equation}
\begin{aligned}
\nabla_{\mathbf{X}}\nabla_{\mathbf{X}'}^{\mathsf{T}}k_{\nu}(\mathbf{X},\mathbf{X}')
={}&-\frac{k_{\nu}'(r)}{r}\,\Lambda^{-1} \\
&-\frac{k_{\nu}''(r)-k_{\nu}'(r)/r}{r^{2}} \\
&\quad\times\Lambda^{-1}\Delta\mathbf{X}\,\Delta\mathbf{X}^{\mathsf{T}}\Lambda^{-1}.
\end{aligned}
\label{eq:matern_mixed}
\end{equation}

The apparent singularities at $r=0$ are removable for $\nu>1$: one evaluates Eqs.~\eqref{eq:matern_grad}--\eqref{eq:matern_mixed} by taking limits as $r\to 0$, yielding $\nabla_{\mathbf{X}}k_\nu=0$ and $\nabla_{\mathbf{X}}\nabla_{\mathbf{X}'}^{\mathsf{T}}k_\nu = -\lim_{r\to 0}\frac{k_\nu'(r)}{r}\,\Lambda^{-1}$.

For $\nu=p+1/2$ with integer $p$, $k_\nu$ reduces to a polynomial in $r$ times $\exp(-\sqrt{2\nu}\,r)$. Two frequently used cases are:
\begin{align}
\nu=\frac{3}{2}: \quad &
k(r)=\sigma_f^2\left(1+\sqrt{3}\,r\right)\exp\!\left(-\sqrt{3}\,r\right), \nonumber\\
&
k'(r)=-3\sigma_f^2\,r\,\exp\!\left(-\sqrt{3}\,r\right),\nonumber\\&
k''(r)=-3\sigma_f^2\left(1-\sqrt{3}\,r\right)\exp\!\left(-\sqrt{3}\,r\right);
\\
\nu=\frac{5}{2}: \quad &
k(r)=\sigma_f^2\left(1+\sqrt{5}\,r+\frac{5}{3}r^{2}\right)\exp\!\left(-\sqrt{5}\,r\right), \nonumber\\
&
k'(r)=-\frac{5}{3}\sigma_f^2\,r\left(1+\sqrt{5}\,r\right)\exp\!\left(-\sqrt{5}\,r\right), \nonumber\\&
k''(r)=\frac{5}{3}\sigma_f^2\left(5r^{2}-\sqrt{5}\,r-1\right)\exp\!\left(-\sqrt{5}\,r\right).
\end{align}
Substituting these $k'(r)$ and $k''(r)$ into Eqs.~\eqref{eq:matern_grad}--\eqref{eq:matern_mixed} yields analytic gradients and mixed second derivatives for the Mat\'ern kernel under ARD length scales.

\bibliographystyle{apsrev4-2}
\bibliography{UEOS}
\end{document}